\documentclass[sn-nature]{sn-jnl}

\usepackage{graphicx}%
\usepackage{multirow}%
\usepackage{amsmath,amssymb,amsfonts}%
\usepackage{amsthm}%
\usepackage{mathrsfs}%
\usepackage[title]{appendix}%
\usepackage{xcolor}%
\usepackage{textcomp}%
\usepackage{manyfoot}%
\usepackage{booktabs}%
\usepackage{algorithm}%
\usepackage{algorithmicx}%
\usepackage{algpseudocode}%
\usepackage{listings}%

\theoremstyle{thmstyleone}%
\theoremstyle{thmstyletwo}%

\theoremstyle{thmstylethree}%

\begin{document}

\title[Article Title]{A Topotactic Phase Transition in the Uranium Oxide System}


\author*[1]{\fnm{Jacek} \sur{Wasik}}\email{jacek.wasik@bristol.ac.uk}

\author[2]{\fnm{Renaud} \sur{Podor}}

\author[1,3]{\fnm{Jarrod} \sur{Lewis}}

\author[1]{\fnm{Niamh} \sur{Cuffe}}

\author[1,4]{\fnm{Connor} \sur{Beer}}

\author[1]{\fnm{Christopher} \sur{Bell}}

\author*[1]{\fnm{Ross} \sur{Springell}}\email{phrss@bristol.ac.uk}


\affil[1]{\orgname{School of Physics}, \orgname{University of Bristol}, \orgaddress{\street{Tyndall Avenue}, \city{Bristol}, \postcode{BS8 1TL}, \country{UK}}}

\affil[2]{\orgname{ICSM}, \orgaddress{\street{Univ Montpellier, CEA, CNRS, ENSCM}, \city{Marcoule}, \postcode{30207}, \country{France}}}

\affil[3]{\orgname{Department of Materials}, \orgname{University of Oxford}, \orgaddress{\street{21 Banbury Road}, \city{Oxford}, \postcode{OX2 6NN}, \country{UK}}}

\affil[4]{\orgname{Department of Materials Design and Manufacturing, School of Engineering}, \orgname{University of Liverpool}, \orgaddress{\street{The Quadrangle}, \city{Liverpool}, \postcode{L69 3GH}, \country{UK}}}


\abstract{A topotactic phase transition involves the transformation of one crystalline solid to another, which may include the loss or gain of material, where the orientation of the parent crystal determines the orientation of the daughter. We set out an experimental approach, based on polyepitaxial thin film deposition, where the precise transformation mechanism in important physico-chemical processes can be revealed in brilliant detail. Here, we find a reversible topotactic transition from (001) cubic UO\textsubscript{2} to a (130) orthorhombic U\textsubscript{3}O\textsubscript{8} structure; a $>$35\% expansion/contraction. This remarkable result solves a puzzle that has eluded researchers for decades, and presents a method for determining the mechanism of crystallographic transformation in many other compounds.}




\maketitle

\section{Introduction}\label{sec1}

The transformation of one crystal phase to another is important for our basic understanding of crystal engineering in the solid state \citep{desiraju2007crystal, nangia2019crystal} and it is crucial to numerous manufacturing processes \citep{kalpakjian2013manufacturing, bhadeshia2017geometry}, materials synthesis routes \citep{callister2020callister}, and degradation mechanisms \citep{roberge2008corrosion, birks2006introduction}. These types of transformations can involve changes in symmetry, variation of bond lengths, bond breaking and formation, and changes in valency/speciation \citep{fultz2020phase, smart2016solid}. A special case of this transformation is that of the topotactic phase transition, where the parent crystal shares a specific orientational relationship with the daughter crystal \citep{meng2023topotactic}. These details are very difficult for theoretical simulations to accurately predict as there are often many energy equivalent possibilities within the uncertainty of the model. Here, we present a method of thin film engineering based on polyepitaxial growth \citep{wasik2024polyepitaxial, wasik2021oxidation}, that can be used to study numerous crystal planes simultaneously, in order to select the single crystal orientation to determine the crystallographic relationship between parent and daughter structure \citep{ wasik2021oxidation}. This method has been applied to the uranium oxide system; a phase diagram rich in complex solid state phenomena \citep{belle1961uranium, mceachern1998review, bevan1986crystal, rousseau2006detailed, leinders2016low, wasik2024polyepitaxial, wasik2021oxidation} and one that has significant industrial relevance, but it could be applied to a host of other important compounds and their transformations \citep{li2019superconductivity, tominaka2014topotactic, meng2023topotactic, kim2020situ, jeen2013topotactic, khare2017topotactic, dasgupta1961topotactic, lee2020aspects, fu2017enhanced, christensen1997phonons, hunter2010topotactic, goto2021synthesis, kloss2021preparation}.

Topotactic transitions involve a structural phase change through the ordered loss, gain, or rearrangement of atoms, without loss of integrity or decomposition \citep{meng2023topotactic}. Not only are they a rich area of fundamental study in their own right, as we present here, but they also offer alternative, lower temperature routes to the traditional solid state methods of material synthesis \citep{li2019superconductivity, abbate1993electronic}. These transitions can be achieved by a wide range of methods, such as simple annealing \citep{kawai2009reversible, hu2022atomic, nomoto2020texture}, redox reactions \citep{lee2018redox, vaney2022topotactic, khare2017topotactic}, electrochemistry \citep{takimoto2023platinum, park2016electrochemical}, electron beam heating \citep{yao2014electron} and proton intercalation \citep{chen2019versatile}, for example. The subtle control of the thermodynamics in these methods also allows the access of metastable states \citep{meng2023topotactic}, often inaccessible via higher temperature chemical synthesis. The recent breakthrough in superconducting infinite-layer nickelates and quintuple-layer square-planar nickelates \citep{li2019superconductivity, pan2022superconductivity}, has heightened interest in this approach for engineering new oxides \citep{meng2023topotactic}. Unlike traditional chemical doping, which often results in random atomic substitutions, topotactic transitions involve long-range ordered structural changes, providing a unique method for creating new oxide materials or phases.

The methodology presented here has been applied to the uranium oxide system, due to a wealth of experience and literature on polycrystal \citep{popel2016structural, wasik2021oxidation, wasik2024polyepitaxial, springell2023review} and single crystal thin film growth \citep{bao2013antiferromagnetism, steeb1961elektronenbeugungs, rennie2018role, wasik2021oxidation}, and due to long-standing questions about the oxidation mechanism of uranium dioxide (UO\textsubscript{2}) \citep{cubicciotti1952reaction}. The oxidation of UO\textsubscript{2} is particularly important because it has been the primary fission fuel used for over 70 years and will likely be needed in some form in future fourth-generation reactors \citep{wang2023nuclear}, with significant efforts dedicated to both experimental methodologies \citep{mceachern1998review, springell2023review} and theoretical models \citep{eriksen2012radiation}. Despite that long history, the detailed mechanism by which UO\textsubscript{2} transforms into U\textsubscript{3}O\textsubscript{8} is still an outstanding puzzle \citep{belle1961uranium, mceachern1998review, bevan1986crystal, rousseau2006detailed, leinders2016low, wasik2024polyepitaxial, wasik2021oxidation, ilton2011xps, spurgeon2019nanoscale}. This transformation is relevant to various aspects of nuclear fuel processing \citep{tomar2023nuclear}, particularly in non-standard operational and spent fuel storage scenarios \citep{international2022status}, where uranium oxides are exposed to a variety of oxidising/reducing conditions. Recently, there has been a significant increase in efforts to study this oxidation process in greater detail \citep{spurgeon2019nanoscale, popel2020atomic, middleburgh2021structure, springell2023review, vallejo2022advances}. Most earlier studies have used powdered or polycrystalline solids \citep{mceachern1998review, idriss2010surface, leinders2016low, rousseau2006detailed}, where surface-to-volume ratio and grain size/grain boundary density consistently affect the results. An alternative approach, using thin films, offers a unique platform for controlling and engineering phase, stoichiometry, and strain \citep{springell2023review}, and where polyepitaxy, in particular, can simplify a three dimensional grain structure into a simpler two-dimensional one \citep{wasik2024polyepitaxial}. There is a wealth of literature on the production of high-quality single crystal thin film samples \citep{springell2023review}, allowing precise control of crystal orientation and strain on a diverse range of readily available substrates, such as yttria-stabilized zirconia (YSZ) \citep{strehle2012characterization}, calcium fluoride (CaF\textsubscript{2}) \citep{bao2013antiferromagnetism}, strontium titanate (SrTiO\textsubscript{3}) \citep{rennie2018study}, and lanthanum aluminate (LAO)  \citep{bao2013antiferromagnetism}. Here, we show epitaxial thin film growth of UO\textsubscript{2} with [100], [110], and [111] orientations normal to the surface of YSZ and CaF\textsubscript{2} substrates. We monitored transformations, \textit{in situ}, using scanning electron microscopy (SEM), X-ray diffraction (XRD), and X-ray reflectivity (XRR).

The current accepted models all regard the \{111\} closest packed UO\textsubscript{2} plane as the most significant and expect transformations to propagate along $<$111$>$ directions, based on oxidation rate data; fastest along $<$111$>$, then $<$110$>$ and slowest along $<$001$>$ \citep{allen1987oxidation}, and based on suggested crystallographic relationships between UO\textsubscript{2} and U\textsubscript{3}O\textsubscript{8} \citep{allen1995mechanism}. We do not believe this to be true, as recent data suggests a different mechanism \citep{wasik2024polyepitaxial} and this paper confirms an alternative route. In this work, we observe a topotactic transformation from \{001\} UO\textsubscript{2} to (130) U\textsubscript{3}O\textsubscript{8}, incorporating a remarkable $>$35\% volume expansion without loss of integrity. Moreover, this transformation is fully reversible. This has direct impact on the way that we understand and model this oxidation process in the nuclear fuel cycle. The methodology presented here could also pave the way for the study of other compounds, where epitaxial films could expose the detailed transformation mechanism as the metal ion valency changes and one crystal topotactically transforms to another.

\begin{figure}[h]%
\centering
\includegraphics[width=\textwidth]{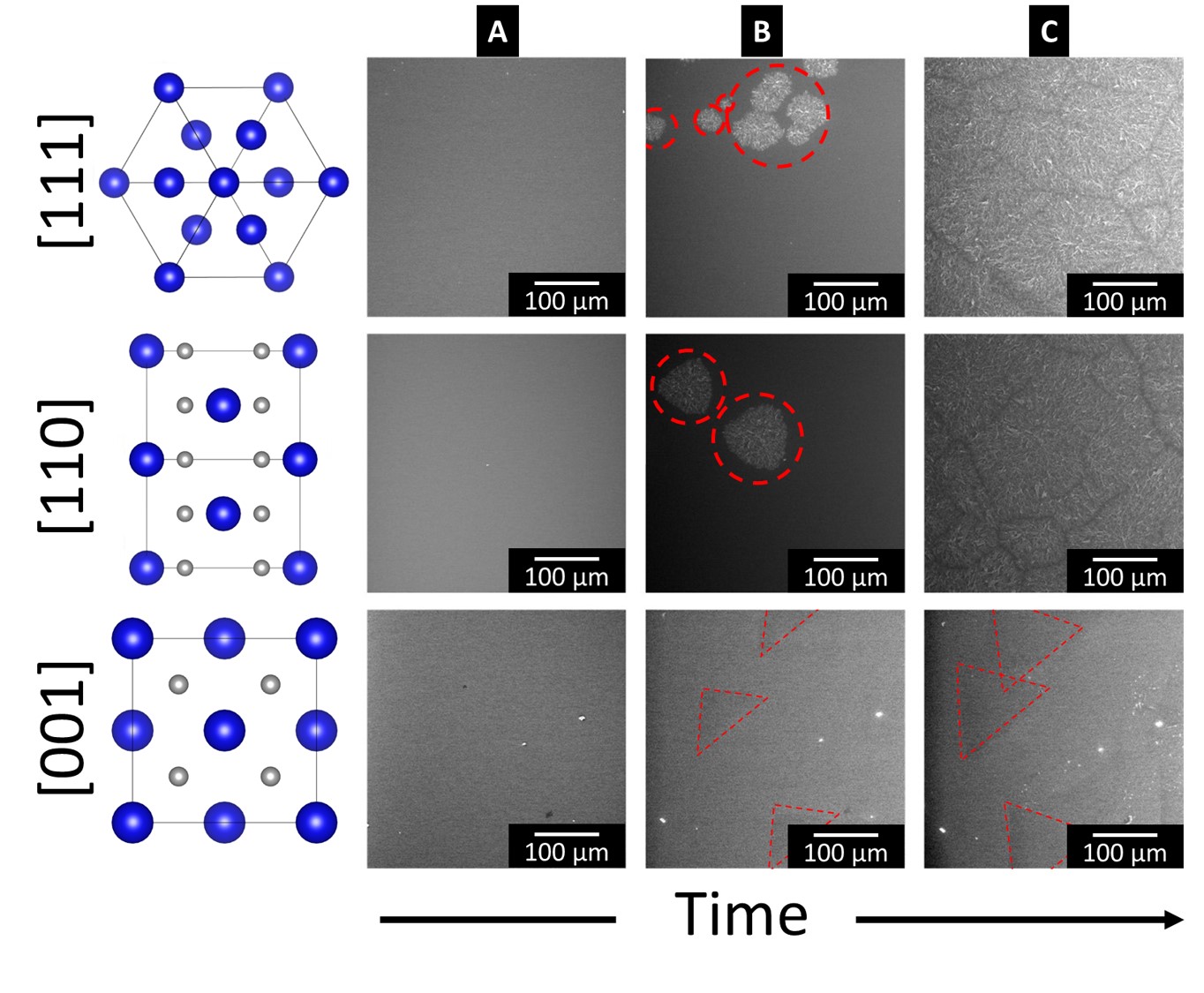}
\caption{\textbf{\textit{In-situ} HT-ESEM oxidation of three principal UO\textsubscript{2} orientations.} From the left, schematic of the three principle cubic orientations of UO\textsubscript{2}, uranium atoms are shown in blue and oxygen atoms in gray. (A) are SEM images of the pristine surface of three samples before oxidation under 3.5 mbar O\textsubscript{2} at 500\textdegree{}C. (B) SEM images shows the same area during the phase transition to U\textsubscript{3}O\textsubscript{8}. Red dashed zones highlight examples of regions of oxidation to U\textsubscript{3}O\textsubscript{8}. Panel (C) shows images taken after full oxidation to U\textsubscript{3}O\textsubscript{8}.}\label{SEM}
\end{figure}


\section{Results}\label{sec2}

Single crystal UO\textsubscript{2} thin film samples representing three principal UO\textsubscript{2} orientations: [111], [110], and [001], deposited on single crystal YSZ substrates, were oxidised \textit{in situ} in a high-temperature environmental scanning electron microscope (HT-ESEM). The HT-ESEM images (Fig.\ref{SEM}) taken at the beginning (A), during (B), and after (C) the oxidation process for these three different single crystal thin films of UO\textsubscript{2} to U\textsubscript{3}O\textsubscript{8} show the characteristic nucleation-and-growth mechanism associated with this oxidative phase transition \citep{mceachern1998review, aronson1957kinetic, hoekstra1961low, westrum1962triuranium, rousseau2006detailed, desgranges2011refinement, walker1965oxidation, allen1995mechanism}. The spallation of U\textsubscript{3}O\textsubscript{8} in the [111] and [110]-oriented single crystals is evidently a result of the 36 \% volume expansion associated with the conversion to the higher oxide \citep{mceachern1998review, taylor1992early, bae1994oxidation}. The previously reported loss of the UO\textsubscript{2} matrix integrity leads to the formation of a popcorn-like morphology for powders \citep{mceachern1998review, quemard2009origin}, and in the case of UO\textsubscript{2} pellets a cauliflower appearance is observed \citep{quemard2009origin, taylor1992early, bae1994oxidation}. In this study, we observe the development of a flake structure, primarily attributed to the thin film nature of the initial material.   

A distinct behaviour is observed for the [001]-oriented epitaxial thin film of UO\textsubscript{2}. The sample does not disintegrate after exposure to oxygen at high temperatures and transforming to U\textsubscript{3}O\textsubscript{8}, with the film instead forming a possible domain structure as shown by the highlighted  triangular regions in Fig.\ref{SEM}. This indicates that the volume expansion of [001]-UO\textsubscript{2}, associated with the formation of U\textsubscript{3}O\textsubscript{8} can be accommodated in the new structure without loss of integrity, which has not been previously reported. Taking these results together, the behaviour contradicts the previously suggested epitaxial relationship between the $<$111$>$ planes of UO\textsubscript{2} and the orthorhombic structure of U\textsubscript{3}O\textsubscript{8}, which was further shown in Fig.\ref{SEM} as no such retention of the thin film integrity when investigating the [111]-oriented sample. \citep{allen1986formation, allen1987oxidation}. The oxidation of the [001] surface of UO\textsubscript{2} deposited on a [001]-YSZ substrate was further studied through a combined \textit{in-situ} x-ray diffraction (XRD) and x-ray reflectivity (XRR) oxidation experiment.

\begin{figure}[h]%
\centering
\includegraphics[width=\textwidth]{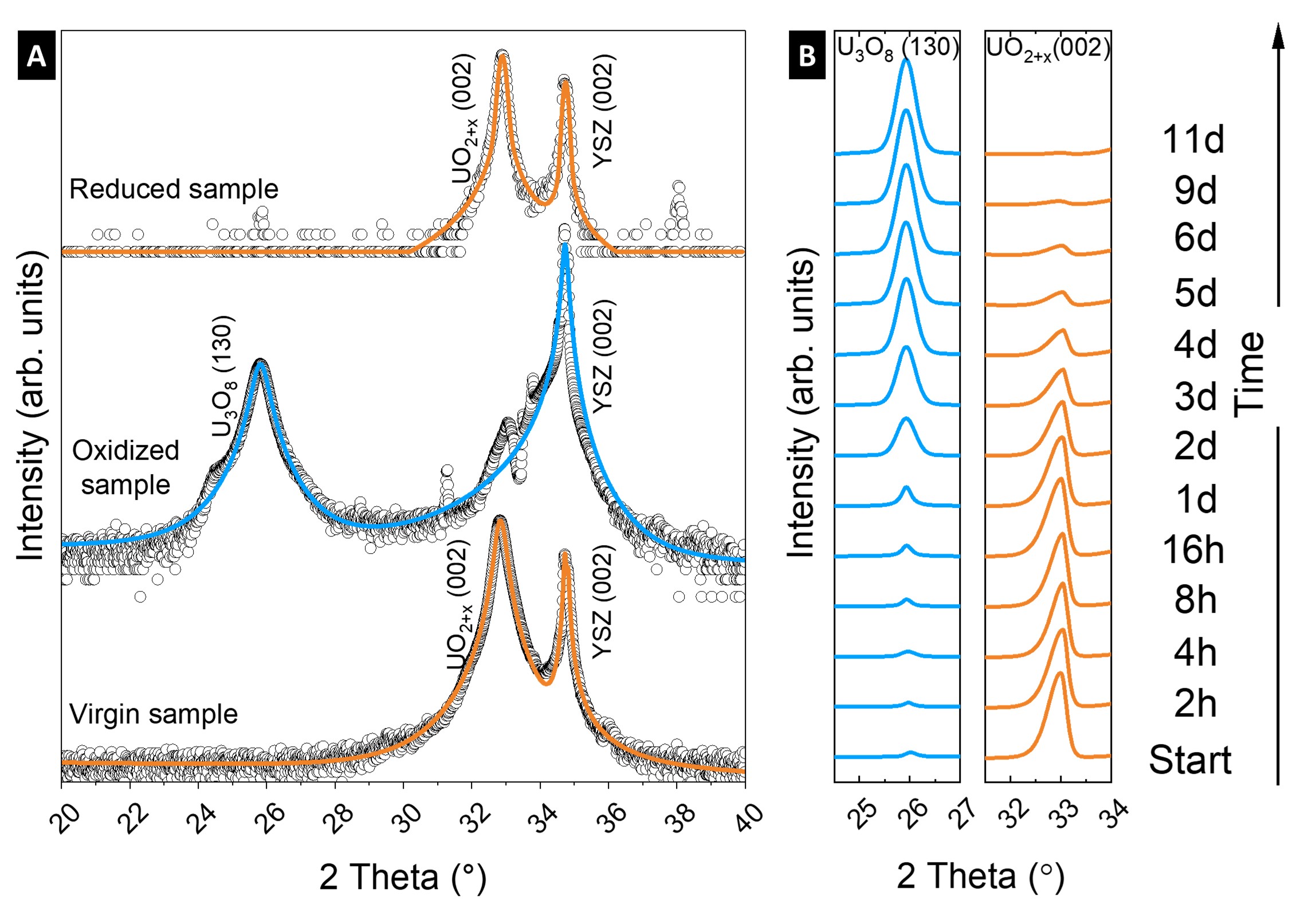}
\caption{\textbf{Reversible topotactic transition between UO\textsubscript{2} and U\textsubscript{3}O\textsubscript{8}.} Figure presenting topotactic phase transition from (001) UO\textsubscript{2} (orange) single crystal thin film deposited onto (001) YSZ to (130) single crystal of U\textsubscript{3}O\textsubscript{8} (blue). (A) shows XRD diffraction pattern from as grown UO\textsubscript{2} (bottom), after oxidation to U\textsubscript{3}O\textsubscript{8} (middle), and reduction back to UO\textsubscript{2} (top), focused on the (002) Bragg peak for cubic UO\textsubscript{2} (orange) and the (130) Bragg peak for orthorhombic U\textsubscript{3}O\textsubscript{8} (blue). The position of the Bragg reflection associated with the YSZ substrate remains unchanged at 34.82\textdegree{}. (B) shows the evolution of the aforementioned Bragg peaks collected during \textit{in-situ} oxidation at 200 mbar of oxygen and 300\textdegree{}C for 11 days. The data plotted in this figure is represented by open circles and the fit envelop by the line.}\label{Topo}
\end{figure}

The XRD data collected from a [001] single crystal thin film before oxidation are shown in Fig. \ref{Topo}A (virgin sample). As shown on a longitudinal scan prior to oxidation (orange bottom data) only two peaks with relatively high intensity are present, corresponding to (002) reflections from UO\textsubscript{2} and YSZ. The central Bragg peak positions were extracted from those fits, which were used to calculate specular lattice parameters for the thin film and substrate, which are 5.467(3)\AA{} and 5.154(2)\AA{}, respectively. This is in good agreement for the data reported with bulk UO\textsubscript{2} \citep{desgranges2009neutron}.

The \textit{in-situ} oxidation experiment was conducted at two different temperatures. In the first step the sample was exposed to 200 mbar of oxygen at 150°C, to allow full oxidation to U\textsubscript{3}O\textsubscript{7} \citep{mceachern1998review, rousseau2006detailed, leinders2016low}. Overnight oxidation for more than 16h did not show any further changes after the initial 2h exposure (Supplementary Fig.1). The temperature was increased to 300°C to promote the phase transition into orthorhombic U\textsubscript{3}O\textsubscript{8}. The subsequent reduction of U\textsubscript{3}O\textsubscript{8} back to UO\textsubscript{2} was performed \textit{ex situ} in a stainless-steel vessel under a controlled hydrogen partial pressure at elevated temperature, following established procedures \citep{pijolat1997reduction}.

The fitting of the high-angle data collected for the (002) Bragg peak of the UO\textsubscript{2+x} and (130) peak of the U\textsubscript{3}O\textsubscript{8} during the experiment is presented in Fig. \ref{Topo}B. At the beginning of the oxidation process a small reflection associated with the (130) Bragg peak for the U\textsubscript{3}O\textsubscript{8} structure appeared, while there was not much change observed at the (002) reflection from the UO\textsubscript{2+x} structure. Significant change to the structure of the sample occurs after two days, when the decrease of the UO\textsubscript{2+x} intensity is noticeable and the signal for the orthorhombic U\textsubscript{3}O\textsubscript{8} becomes stronger.

The high-angle scan of the oxidised sample, measured at room temperature, is shown in Fig. \ref{Topo}A (Oxidized sample). Utilizing the central Bragg peak position obtained from the fit to the data, the lattice spacing for (130) U\textsubscript{3}O\textsubscript{8} was calculated as 3.428(3)\AA{}. The measured value aligns well with the reported bulk material value of 3.429\AA{} \citep{ackermann1977thermal}. The Bragg reflection position, associated with the YSZ substrate, remained unchanged with the same lattice parameter of 5.154(2)\AA{}. It is also important to note that both UO$_{2}$ and U$_{3}$O$_{8}$ phases coexist, and that one is gradually replacing the other during the oxidation process. This implies, together with the maintained integrity of the films observed via SEM, that the U$_{3}$O$_{8}$ forms initially at the surface of the UO$_{2}$, and the transformation proceeds along the growth direction of the crystal. Therefore this topotactic transition should be observed in any system that promotes this growth axis, i.e. it is substrate-independent in that sense (indeed we have also observed this phenomenon on CaF$_{2}$).

The high-angle XRD scan of the sample reduced from U\textsubscript{3}O\textsubscript{8} back to UO\textsubscript{2} is shown in Fig. \ref{Topo}A. The lattice parameter derived from the central Bragg peak position was determined to be 5.447(3) Å, which is slightly smaller than the initial value, but still within values reported for  bulk UO\textsubscript{2} \citep{desgranges2009neutron}. This lattice contraction may arise from residual hyper-stoichiometry, and compressive strain at the UO\textsubscript{2}/YSZ interface, both of which are known to decrease the lattice spacing \citep{willis1978defect, strehle2012characterization, springell2023review}.

\begin{figure}[h]%
\centering
\includegraphics[width=\textwidth]{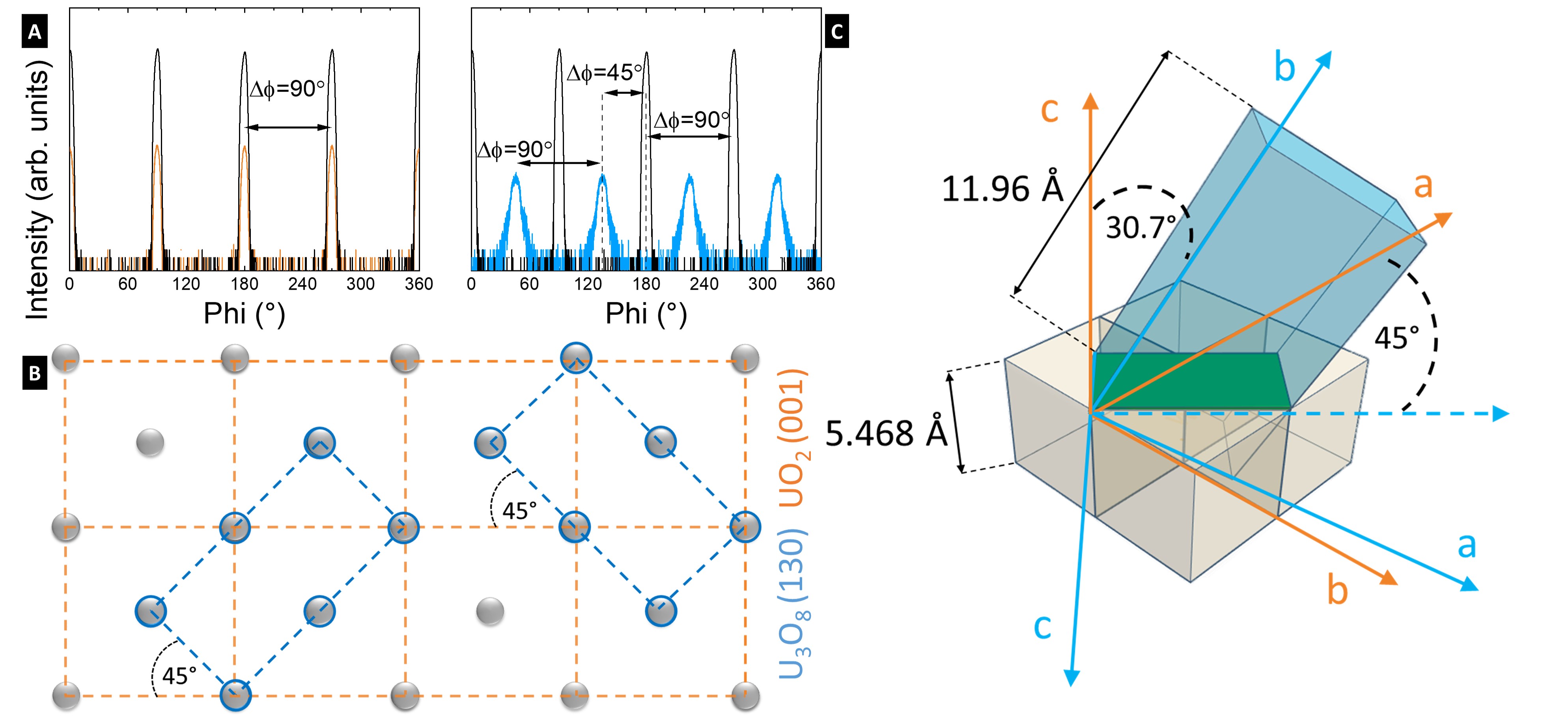}
\caption{\textbf{Establishing the orientation relationship of the topotactic transformation.} Figure presenting relationship between UO\textsubscript{2} (orange) single crystal thin and (130) single crystal of U\textsubscript{3}O\textsubscript{8} (blue). (A) shows the (024) off-specular UO\textsubscript{2} (orange) reflection on (024) off-specular YSZ (black), and the (261) reflection from U\textsubscript{3}O\textsubscript{8} (blue). A schematic representation of the planes arrangement between cubic UO\textsubscript{2}/YSZ and two arrangements of U\textsubscript{3}O\textsubscript{8} is shown on panel (B). The 3D representation of epitaxial relation is illustrated on panel (C).}\label{Topo2}
\end{figure}

The orientational relationships between the films and substrate were determined by off-specular XRD scans. Fig \ref{Topo2}A shows $\varphi$-scans of the YSZ substrate (black), the initial UO\textsubscript{2} sample (orange), and the final U\textsubscript{3}O\textsubscript{8} film, confirming the single crystal nature of the samples. The UO\textsubscript{2} relationship to the substrate is cube on cube, and is well-known in literature \citep{bao2013antiferromagnetism, rennie2018role, springell2023review}. The $\varphi$-scan of U\textsubscript{3}O\textsubscript{8} (Fig. \ref{Topo2}A blue) corresponds to the epitaxial relationship between the (00l) YSZ and the (130) U\textsubscript{3}O\textsubscript{8}, exhibiting a separation angle between the film and the substrate to be $\Delta\phi=45$\textdegree{}. This indicates the existence of two domains of U\textsubscript{3}O\textsubscript{8} with $\Delta\phi=90$\textdegree{} rotation.

The schematic of the epitaxial relationship between the (001) plane of UO\textsubscript{2} and the (130) planes of U\textsubscript{3}O\textsubscript{8} is illustrated in Fig \ref{Topo2}B. The three-dimensional representation of this model is shown of Fig \ref{Topo2}C. While the off-axis XRD scans were conducted after the full oxidation into U\textsubscript{3}O\textsubscript{8} and exhibit a crystallographic relation with YSZ, the applicability of this model extends to UO\textsubscript{2} due to the analogous structures (epitaxial match) of UO\textsubscript{2} and YSZ. This picture is supported by the concurrent presence of both oxides throughout the oxidation experiment (Fig. \ref{Topo}B). For UO\textsubscript{2} with lattice spacing of 5.47\AA{}, the strain between UO\textsubscript{2} and U\textsubscript{3}O\textsubscript{8} on the shorter axis is 1.07\%, while on the longer axis, it is 7.26\%. These values indicate a higher degree of preference compared to the YSZ structure, where the corresponding strains would be 7.36\% and 13.93\%, respectively. This transition has also been observed for (001) UO\textsubscript{2} deposited onto (001) CaF\textsubscript{2} (Supplementary Fig.2). Furthermore, it is plausible that this model holds true for other epitaxial substrates of UO\textsubscript{2}.

\begin{figure}[h]%
\centering
\includegraphics[width=\textwidth]{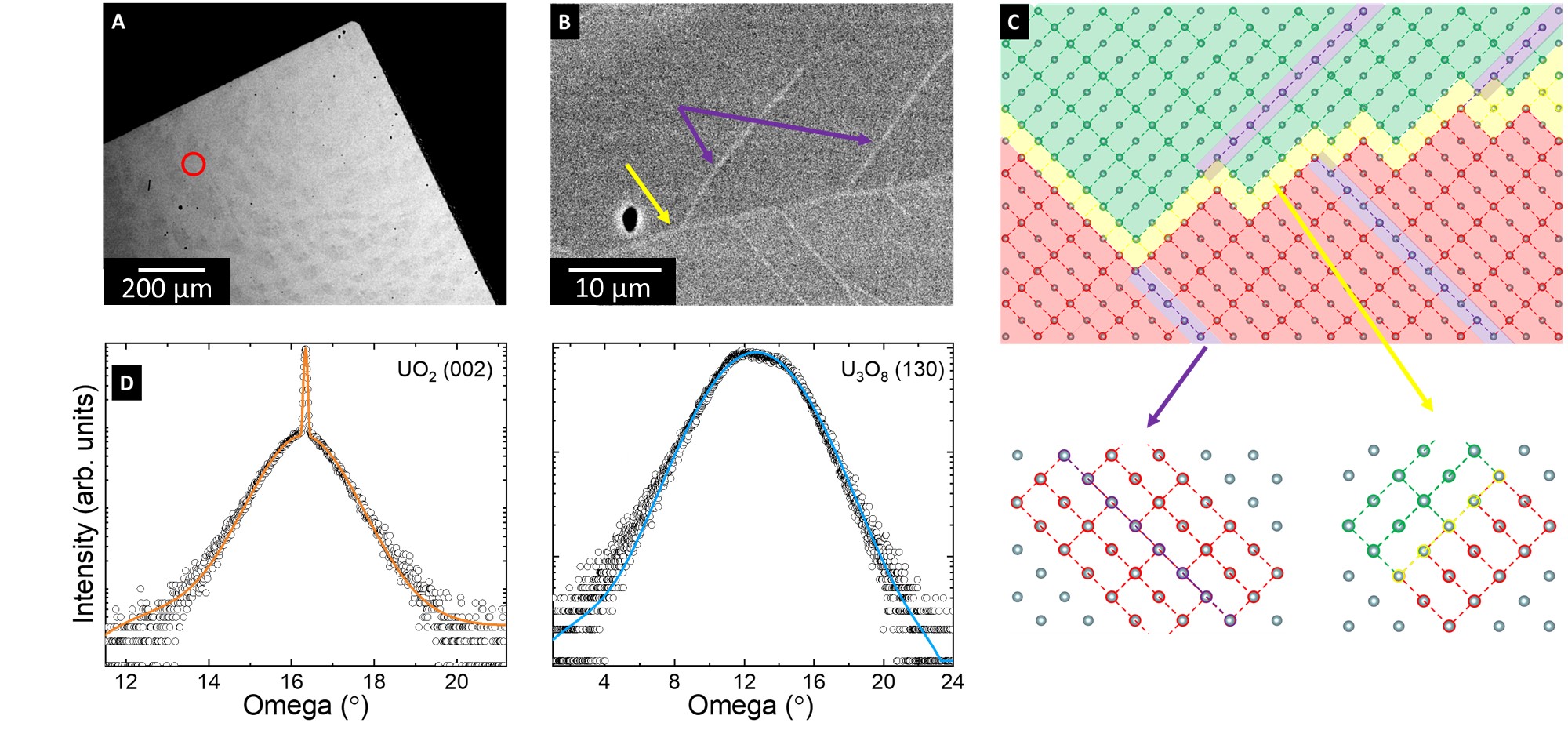}
\caption{\textbf{Assessing domain formation and resulting defects in the UO\textsubscript{2} to U\textsubscript{3}O\textsubscript{8} topotactic oxidation.} (A) shows SEM image of oxidised sample at low magnification, with multiple domains visible. Magnification of an area between two domains is shown on panel (B), where two types of boundaries can be observed. The two types of the boundaries are schematically shown (depicted uranium atoms only) on panel (C) where one is formed within the same domain by shifting U\textsubscript{3}O\textsubscript{8} by one unit cell of UO\textsubscript{2} (purple), and the second one is formed between two perpendicular domains of U\textsubscript{3}O\textsubscript{8} (yellow). (D) The formation of the domains and resulting increase of defects can be observed in rocking curve measurements performed before oxidation on (002) Bragg reflection from UO\textsubscript{2} (orange fit envelop) and after oxidation from (130) Bragg peak from U\textsubscript{3}O\textsubscript{8} (blue fit envelop).} \label{defects}
\end{figure}

The crystalline quality of the sample was assessed pre- and post-oxidation by performing rocking curve scans, from which the mosaicity of the thin film can be quantified. The data collected prior to the oxidation experiment, of the (002) UO\textsubscript{2} Bragg peak is shown in Fig. \ref{defects}D (orange data). To fit the profile (orange data) two functions were used, Gaussian for the narrow component and Pearson VII for the broad part. The FWHM for the sharp part is 0.072$\pm$0.002\textdegree{} and for the wide part is 1.52$\pm$0.02\textdegree{}. The rocking curve profile for the (130) Bragg peak of U\textsubscript{3}O\textsubscript{8} after oxidation is shown at Fig. \ref{defects}D (blue). To fit this data a Pearson VII function was used, giving the FWHM of 4.91$\pm$0.03\textdegree{}, with no sharp feature present.

The change in the rocking curve from a narrow FWHM observed for (002) UO\textsubscript{2} Bragg peak to a very broad profile of the (130) U\textsubscript{3}O\textsubscript{8} Bragg peak indicates an increase in mosaicity. This might be connected with the formation of larger domains observed in the HT-ESEM experiment (Fig. \ref{SEM}), and was further investigated by SEM on the oxidised sample. Fig. \ref{defects}A displays a low-magnification back-scattered SEM image of the transformed sample, revealing the presence of multiple domains. Zooming in on the region between two domains (Fig \ref{defects}B) shows the  potential formation  of two distinct types of defects.

The observed epitaxial relationship between cubic UO\textsubscript{2} single crystal and layered orthorhombic structure of U\textsubscript{3}O\textsubscript{8} allows for formation of two domains. Two possible arrangements of the (130) plane of the U\textsubscript{3}O\textsubscript{8} structure on the (001) plane of UO\textsubscript{2} (Fig. \ref{Topo2}B), are indicated by the red and green dashed lines in Fig. \ref{defects}C. As a consequence of this model, two different boundaries within the U\textsubscript{3}O\textsubscript{8} would be possible. The first type of boundary, shown with yellow colour in Fig. \ref{defects}C, would be formed between two domains where unit cells of U\textsubscript{3}O\textsubscript{8} are rotated by 90\textdegree{} with respect to each other. The second type would be a defect within the same domain, where unit cells are parallel to each other but shifted by a half unit cell along the longer axis, shown in Fig. \ref{defects}C in purple. This model would also offer an explanation for the two defect types observed in Fig. \ref{defects}B. The presence of such domains these two types of defects in a single crystal structure would lead to a broader rocking curve due to a higher mosaic spread. Therefore, this explains the broadening of the rocking curve after oxidation of UO\textsubscript{2} to U\textsubscript{3}O\textsubscript{8} shown of Fig. \ref{defects}D.

\section{Discussion}\label{sec121}

Recent studies have significantly advanced our understanding of the oxidation behaviour of UO\textsubscript{2} thin films, challenging long-held assumptions \citep{wasik2024polyepitaxial}. Allen \textit{et al.} \citep{allen1986formation, allen1987oxidation} hypothesized a transformation of (111) UO\textsubscript{2} planes into the (001) plane of U\textsubscript{3}O\textsubscript{8} based on shear mechanisms that allow planes to change their (ABC) stacking to (A), which was thought to explain the accelerated oxidation of \{111\} UO\textsubscript{2}. However, recent findings have shown this hypothesis to be inaccurate, with the fastest oxidation rate observed for [001] UO\textsubscript{2} \citep{wasik2024polyepitaxial}. Additionally, the previously suggested absence of an epitaxial relationship involving the (001) plane of UO\textsubscript{2}, which was believed to hinder oxidation due to lattice expansion constraints, is not supported by the present results.

Topotactic transitions in uranium oxides have previously been reported by Desgranges \textit{et al.} \citep{desgranges2010influence} during the early oxidation of UO\textsubscript{2} to U\textsubscript{3}O\textsubscript{7}. In that work, the fluorite cubic lattice transforms epitaxially into a slightly distorted tetragonal structure ($c/a \approx 1.03$), preserving crystallographic orientation across the interface. However, the formation of alternating domains introduces localized stress concentrations, ultimately leading to cracking beyond a critical thickness.

By contrast, the present study reveals a significantly more extensive structural transformation. The oxidation proceeds from the cubic UO\textsubscript{2}–U\textsubscript{4}O\textsubscript{9} phase to orthorhombic U\textsubscript{3}O\textsubscript{8}, accompanied by a volumetric expansion of approximately 36\% and a fundamental change in crystal symmetry. This transformation follows the well-established oxidation sequence UO\textsubscript{2} $\rightarrow$ UO\textsubscript{2+x} $\rightarrow$ U\textsubscript{4}O\textsubscript{9} $\rightarrow$ U\textsubscript{3}O\textsubscript{8} \citep{desgranges2009neutron}, reflecting progressive oxygen incorporation and the evolution of intermediate phases. Importantly, the process is reversible under suitable reduction conditions \citep{pijolat1997reduction}. There was no obvious loss in structural integrity during the forward and reverse processes, however, further studies would be required to investigate the possible accumulation of defects as this transformation is cycled.

The HT-ESEM observations confirm a nucleation and growth mechanism for all orientations. However, a strong crystallographic dependence is observed. The [111] and [110] orientations exhibit spallation and loss of integrity due to the large volume expansion, consistent with previous studies. In contrast, the [001] oriented thin film maintains structural integrity throughout oxidation, forming a domain structure that accommodates the strain without fragmentation. This indicates that the transformation strain can be effectively accommodated along this orientation.

This crystallographically dependent oxidation behaviour differs from trends typically observed for dissolution rates \citep{rennie2018role}. Although corrosion behaviour is strongly influenced by oxidation processes, dissolution is ultimately governed by surface bond breaking and detachment kinetics. Consequently, even if the topotactic transition accelerates oxidation, it may simultaneously preserve surface integrity by limiting spallation and mechanical fragmentation, thereby reducing the effective dissolution rate. As a result, crystallographic orientations that facilitate oxygen uptake and transport can exhibit enhanced oxidation kinetics without necessarily corresponding to higher dissolution rates.

The topotactic phase transition from (001) UO\textsubscript{2} to (130) U\textsubscript{3}O\textsubscript{8} is confirmed by XRD analysis. Off-specular $\varphi$-scans demonstrate a clear epitaxial relationship, with a separation angle of $\Delta\phi=45$\textdegree{} and the presence of two domains rotated by $\Delta\phi=90$\textdegree{}. Strain analysis indicates stronger lattice compatibility at the UO\textsubscript{2+x}–U\textsubscript{3}O\textsubscript{8} interface than at the U\textsubscript{3}O\textsubscript{8}–YSZ interface, suggesting that the transformation is primarily governed by the UO\textsubscript{2+x} lattice rather than the substrate. This conclusion is supported by experiments on CaF\textsubscript{2}, where the same transformation is observed. Furthermore, the coexistence of parent and product phases during oxidation confirms that the transformation progresses along the growth direction, consistent with a topotactic mechanism that is largely substrate independent.

Post-oxidation analysis reveals increased mosaicity and the formation of multiple domains, consistent with the broadening observed in rocking curve measurements. SEM imaging identifies two distinct types of domain boundaries, which correspond well with the proposed structural model and explain the observed defect structure.

The preferential formation of the (130) U\textsubscript{3}O\textsubscript{8} orientation is consistent with earlier observations by Song \textit{et al}. \citep{song1994formation}, who reported enhanced (130) and (260) reflections during high-temperature oxidation of UO\textsubscript{2}. While this was previously attributed to strain accommodation, the present work demonstrates that this preference arises from a well-defined epitaxial relationship.

An important aspect of the observed transformation is its reversibility, demonstrating that large structural rearrangements can proceed without loss of crystallinity. This highlights the broader potential of topotactic transitions for engineering functional materials, where reversible changes in crystal structure and electronic properties can be exploited. Such behaviour is particularly relevant for oxide electronics, including resistive switching devices, sensors, and thin film components in MOSFET architectures, where stability under redox cycling is essential.

Finally, the approach presented here, particularly when combined with the previously demonstrated polyepitaxial methodology \citep{wasik2024polyepitaxial}, provides a powerful tool for investigating complex solid state transformations. By enabling the simultaneous study of multiple crystallographic orientations, it allows detailed insight into phase transformations even in extensively studied materials. Its application to uranium oxides demonstrates that significant new understanding can still be achieved in systems that have been investigated for decades, opening pathways for the discovery and control of new phases in complex oxide materials.

In conclusion, we have demonstrated a reversible topotactic transformation of epitaxial UO\textsubscript{2} films into single crystal U\textsubscript{3}O\textsubscript{8}, with the ability to accommodate a 36\% volume expansion without loss of integrity. The identification of the epitaxial relationship, domain formation, and defect structure provides new insight into the oxidation mechanism and establishes a framework for studying similar transformations in other material systems.

\section{Methods}\label{sec11}

\subsection{Reactive DC magnetron sputtering}\label{subsec2}

Epitaxial UO\textsubscript{2} thin film samples were fabricated using reactive DC magnetron sputtering system at the University of Bristol \citep{springell2023review}. The system operates at base pressure of 1 $\times$ 10\textsuperscript{-10} mbar, with 5.5N argon used as the main sputtering gas at the pressure of 7.3 $\times$ 10\textsuperscript{-3} mbar. To grow single crystal UO\textsubscript{2}, a depleted uranium target was used, and a partial pressure 2 $\times$ 10\textsuperscript{-5} mbar of O\textsubscript{2} was applied. To assist with the assembly of a high quality crystal structure, substrates were held at an elevated temperature of 800\textdegree{}C. Yttria stabilized zirconia (YSZ) and CaF\textsubscript{2} substrates obtained commercially from MTI Corporation, with dimensions   $10\,\text{mm}\times10\,\text{mm}\times0.5\,\text{mm}$, polished to 2-3\AA{} root mean square (RMS) roughness were utilized. Both YSZ and CaF\textsubscript{2} provide a good 1:1 epitaxial lattice match with UO\textsubscript{2}, facilitating growth of the single crystal thin films. With calibrated deposition rate, the growth time was controlled to achieve $\sim$ 60 nm thick layer of UO\textsubscript{2}.

\subsection{HT-ESEM}\label{subsec2}

The microscope used in this experiment was a HT-ESEM at the Institut de Chimie Separative de Marcoule (ICSM) in Marcoule, France, model: FEI Quanta 200 FEG ESEM, Thermo Fisher Scientific, Massachusetts, USA. It is a scanning electron microscope fitted with a field emission gun electron source to provide high imaging resolution. This microscope operates in different modes including environmental mode that allows to use air, pure O\textsubscript{2} or a mixture of O\textsubscript{2} and N\textsubscript{2}. The pressure range inside the ESEM chamber can be in range between 10-750 Pa. The sample size is restricted by a 5mm circle. The furnace can heat the sample up to 1200\textdegree{}C and be investigated under magnification between $\times$130 to $\times$100 000.

Samples were stored in air in separate membrane boxes and transported to ICSM in France. The insertion was performed in air. The samples were then heated under an oxygen atmosphere at 350 Pa. The temperature was gradually increased from room temperature to approximately 500 \textdegree{}C, at which point the first signs of oxidation were observed. Once oxidation initiated, the temperature was held constant and not further adjusted throughout the experiment. This oxidation temperature was higher than that used in the XRD measurements, due to the lower oxygen partial pressure and the limited time available on the experimental setup. Throughout the experiment, the sample regions were continuously monitored, with images recorded every 3 to 5 seconds. All images were acquired at the same magnification of $\times$250, corresponding to an area of 512 $\mu$m $\times$ 470 $\mu$m.

The Fiji ImageJ software \citep{schneider2012nih} was used to analyze surface changes during sample treatment. Surface cracking due to oxidation and the expansion of these cracks were monitored by measuring the percentage of the damaged surface relative to the entire ROI. The background of each photo was created using a Gaussian-blurred version of the image, which was then subtracted from the original image. To enhance the contrast between oxidized and pristine areas, a series of edge-finding and smoothing functions were applied. Next, Gaussian-blurred or median filters were used. Finally, areas were identified by selecting a threshold level in the image contrast, with the pristine region falling below this threshold.

\subsection{X-ray analysis}\label{subsec2}

X-ray diffraction (XRD) measurements were performed using a Philips X'Pert PRO MPD diffractometer with a Cu-K$\alpha$ source and beamline I07 at Diamond Light Source, Didcot, UK, with an X-ray beam (8 keV; $\lambda$ = 1.5498 \AA). The \textit{in-situ} oxidation experiments was performed utilizing an Anton Paar HTK 1200 high temperature environment chamber for the diffractometer and a hot stage controlled by a 24V power supply.

The peak fitting for XRD data has been carried out using Line-Profile Analysis Software (LIPRAS) \citep{esteves2017lipras} software. LIPRAS allows for least-squares fitting of Bragg peaks in diffraction data using a graphical user interface. Full description of the errors for all profile parameters were generated by conducting Bayesian inference analysis on least-squares results using a Markov Chain Monte Carlo algorithm.

\backmatter

{\section*{Data Availability}

The data will be available in the University's Data Repository in a form suitable for long-term retention and wider publication. Available from the corresponding author upon request.}

{\section*{Acknowledgements}

This research was supported by the Bristol Centre for Functional Nanomaterials, Centre for Doctoral Training, the University of Bristol Facility for Radioactive Materials Surfaces (FaRMS), funded by the UK Engineering and Physical Sciences Research Council [EP/V035495/1], and the TRANSCEND consortium on nuclear waste and decommissioning, also funded by the UK Engineering and Physical Sciences Research Council [EP/S01019X/1]. We acknowledge the beamline I07 at Diamond Light Source (UK) for beam time (experiment no. SI34673) and the staff there for their help during the experiment.}

{\section*{Author Contributions}

JW -  Conceptualisation, sample fabrication, \textit{in-situ} XRD experiment and data analysis, SEM data collection and analysis, HT-ESEM data analysis, figures, original draft preparation; 
RP - \textit{In-situ} HT-ESEM data collection and analysis; 
JL - sample fabrication, XRD experiments and data analysis; 
NC - XRD experiments and data analysis; 
CB - XRD experiments and data analysis; 
RS - Conceptualisation, supervision, manuscript review and editing; 
All authors discussed and contributed to the writing of the paper.}

{\section*{Competing Interests}

The authors declare no competing interests.}





\bibliography{sn-bibliography}

@article{desgranges2010influence,
  title={Influence of the \uppercase{U\textsubscript{3}O\textsubscript{7}} domain structure on cracking during the oxidation of \uppercase{UO\textsubscript{2}}},
  author={Desgranges, Lionel and Palancher, Herv{\'e} and Gamal{\'e}ri, M and Micha, Jean-S{\'e}bastien and Optasanu, Virgil and Raceanu, Laura and Montesin, Tony and Creton, Nicolas},
  journal={Journal of Nuclear Materials},
  volume={402},
  number={2-3},
  pages={167--172},
  year={2010},
  publisher={Elsevier}
}

@article{wasik2024polyepitaxial,
  title={Polyepitaxial grain matching to study the oxidation of uranium dioxide},
  author={Wasik, Jacek and Sutcliffe, Joseph and Podor, Renaud and Lewis, Jarrod and Darnbrough, James Edward and Rennie, Sophie and Akbar Hussain, Syed and Bell, Christopher and Chaney, Daniel Alexander and Griffiths, Gareth and others},
  journal={npj Materials Degradation},
  volume={8},
  number={1},
  pages={68},
  year={2024},
  publisher={Nature Publishing Group UK London}
}

@article{pan2022superconductivity,
  title={Superconductivity in a quintuple-layer square-planar nickelate},
  author={Pan, Grace A and Ferenc Segedin, Dan and LaBollita, Harrison and Song, Qi and Nica, Emilian M and Goodge, Berit H and Pierce, Andrew T and Doyle, Spencer and Novakov, Steve and C{\'o}rdova Carrizales, Denisse and others},
  journal={Nature materials},
  volume={21},
  number={2},
  pages={160--164},
  year={2022},
  publisher={Nature Publishing Group UK London}
}

@article{lee2020aspects,
  title={Aspects of the synthesis of thin film superconducting infinite-layer nickelates},
  author={Lee, Kyuho and Goodge, Berit H and Li, Danfeng and Osada, Motoki and Wang, Bai Yang and Cui, Yi and Kourkoutis, Lena F and Hwang, Harold Y},
  journal={APL Materials},
  volume={8},
  number={4},
  pages={4},
  year={2020},
  publisher={AIP Publishing}
}

@article{abbate1993electronic,
  title={Electronic structure and spin-state transition of \uppercase{L\lowercase{a}C\lowercase{o}O\textsubscript{3}}},
  author={Abbate, M and Fuggle, JC and Fujimori, A and Tjeng, LH and Chen, CT and Potze, R and Sawatzky, GA and Eisaki, H and Uchida, S},
  journal={Physical Review B},
  volume={47},
  number={24},
  pages={16124},
  year={1993},
  publisher={APS}
}

@article{li2019superconductivity,
  title={Superconductivity in an infinite-layer nickelate},
  author={Li, Danfeng and Lee, Kyuho and Wang, Bai Yang and Osada, Motoki and Crossley, Samuel and Lee, Hye Ryoung and Cui, Yi and Hikita, Yasuyuki and Hwang, Harold Y},
  journal={Nature},
  volume={572},
  number={7771},
  pages={624--627},
  year={2019},
  publisher={Nature Publishing Group UK London}
}

@book{kalpakjian2013manufacturing,
  title={Manufacturing Engineering \& Technology},
  author={Kalpakjian, S. and Schmid, S.},
  isbn={9780133151213},
  year={2013},
  publisher={Pearson Education}
}

@book{callister2020callister,
  title={Callister's Materials Science and Engineering},
  author={Callister, W.D. and Rethwisch, D.G.},
  isbn={9781119453918},
  year={2020},
  publisher={Wiley}
}

@book{roberge2008corrosion,
  title={Corrosion Engineering: Principles and Practice},
  author={Roberge, P.R.},
  isbn={9780071604758},
  year={2008},
  publisher={McGraw-Hill}
}

@article{kim2020situ,
  title={In situ observations of topotactic phase transitions in a ferrite memristor},
  author={Kim, Hyoung Gyun and Nallagatla, Ventaka Raveendra and Kwon, Deok-Hwang and Jung, Chang Uk and Kim, Miyoung},
  journal={Journal of Applied Physics},
  volume={128},
  number={7},
  year={2020},
  publisher={AIP Publishing}
}

@article{jeen2013topotactic,
  title={Topotactic phase transformation of the brownmillerite \uppercase{S\lowercase{r}C\lowercase{o}O\textsubscript{2.5}} to the perovskite \uppercase{S\lowercase{r}C\lowercase{o}O\textsubscript{3-$\delta$}}},
  author={Jeen, Hyoungjeen and Choi, Woo Seok and Freeland, John W and Ohta, Hiromichi and Jung, Chang Uk and Lee, Ho Nyung},
  journal={Advanced Materials},
  volume={25},
  number={27},
  year={2013},
  publisher={Oak Ridge National Lab.(ORNL), Oak Ridge, TN (United States)}
}

@article{tominaka2014topotactic,
  title={Topotactic reduction of oxide nanomaterials: unique structure and electronic properties of reduced \uppercase{T\lowercase{i}O\textsubscript{2}} nanoparticles},
  author={Tominaka, Satoshi and Yoshikawa, Hideki and Matsushita, Yoshitaka and Cheetham, Anthony K},
  journal={Materials Horizons},
  volume={1},
  number={1},
  pages={106--110},
  year={2014},
  publisher={Royal Society of Chemistry}
}

@article{kloss2021preparation,
  title={Preparation of bulk-phase nitride perovskite \uppercase{L\lowercase{a}R\lowercase{e}N\textsubscript{3}} and topotactic reduction to \uppercase{L\lowercase{a}N\lowercase{i}N\textsubscript{2}}-type \uppercase{L\lowercase{a}R\lowercase{e}N\textsubscript{2}}},
  author={Klo{\ss}, Simon D and Weidemann, Martin L and Attfield, J Paul},
  journal={Angewandte Chemie International Edition},
  volume={60},
  number={41},
  pages={22260--22264},
  year={2021},
  publisher={Wiley Online Library}
}

@article{goto2021synthesis,
  title={Synthesis and magnetic properties of tetragonally ordered \uppercase{F\lowercase{e}\textsubscript{2}N\lowercase{i}\textsubscript{2}N} alloy using topotactic nitriding reaction},
  author={Goto, Sho and Kura, Hiroaki and Tsujikawa, Masahito and Shirai, Masafumi and Ito, Keita and Suemasu, Takashi and Takanashi, Koki and Yanagihara, Hideto},
  journal={Journal of Alloys and Compounds},
  volume={885},
  pages={161122},
  year={2021},
  publisher={Elsevier}
}

@article{hunter2010topotactic,
  title={Topotactic nitrogen transfer: structural transformation in cobalt molybdenum nitrides},
  author={Hunter, Stuart M and Mckay, David and Smith, Ronald I and Hargreaves, Justin SJ and Gregory, Duncan H},
  journal={Chemistry of Materials},
  volume={22},
  number={9},
  pages={2898--2907},
  year={2010},
  publisher={ACS Publications}
}

@article{fu2017enhanced,
  title={Enhanced Photoactivity from Single-Crystalline \uppercase{S\lowercase{r}T\lowercase{a}O\textsubscript{2}N} Nanoplates Synthesized by Topotactic Nitridation},
  author={Fu, Jie and Skrabalak, Sara E},
  journal={Angewandte Chemie},
  volume={129},
  number={45},
  pages={14357--14361},
  year={2017},
  publisher={Wiley Online Library}
}

@book{christensen1997phonons,
  title={Phonons and phase transitions in GaN},
  author={Christensen, NE and Perlin, P},
  booktitle={Semiconductors and Semimetals},
  volume={50},
  pages={409--429},
  year={1997},
  publisher={Elsevier}
}

@article{dasgupta1961topotactic,
  title={Topotactic transformations in iron oxides and oxyhydroxides},
  author={Dasgupta, DR},
  journal={Indian Journal of Physics},
  volume={35},
  pages={401--419},
  year={1961}
}

@article{khare2017topotactic,
  title={Topotactic metal--insulator transition in epitaxial \uppercase{S\lowercase{r}F\lowercase{e}O\textsubscript{x}} thin films},
  author={Khare, Amit and Shin, Dongwon and Yoo, Tae Sup and Kim, Minu and Kang, Tae Dong and Lee, Jaekwang and Roh, Seulki and Jung, In-Ho and Hwang, Jungseek and Kim, Sung Wng and others},
  journal={Advanced Materials},
  volume={29},
  number={37},
  pages={1606566},
  year={2017},
  publisher={Wiley Online Library}
}

@book{smart2016solid,
  title={Solid State Chemistry: An Introduction, Fourth Edition},
  author={Smart, L.E. and Moore, E.A.},
  isbn={9781439847923},
  year={2016},
  publisher={CRC Press}
}

@book{fultz2020phase,
  title={Phase Transitions in Materials},
  author={Fultz, B.},
  isbn={9781108622554},
  year={2020},
  publisher={Cambridge University Press}
}

@article{kawai2009reversible,
  title={Reversible changes of epitaxial thin films from perovskite \uppercase{L\lowercase{a}N\lowercase{i}O\textsubscript{3}} to infinite-layer structure \uppercase{L\lowercase{a}N\lowercase{i}O\textsubscript{2}}},
  author={Kawai, Masanori and Inoue, Satoru and Mizumaki, Masaichiro and Kawamura, Naomi and Ichikawa, Noriya and Shimakawa, Yuichi},
  journal={Applied Physics Letters},
  volume={94},
  number={8},
  year={2009},
  publisher={AIP Publishing}
}

@article{nomoto2020texture,
  title={Texture and phase control of magnetron-sputtered \uppercase{VO\textsubscript{2}} thin films with an \uppercase{A\lowercase{l}}-doped \uppercase{Z\lowercase{n}O} seed layer using topotactic oxidization},
  author={Nomoto, Junichi and Yamaguchi, Iwao and Nakajima, Tomohiko and Tsuchiya, Tetsuo},
  journal={Surface and Coatings Technology},
  volume={394},
  pages={125769},
  year={2020},
  publisher={Elsevier}
}

@article{chen2019versatile,
  title={Versatile and highly efficient controls of reversible topotactic metal--insulator transitions through proton intercalation},
  author={Chen, Shanquan and Zhou, Haiping and Ye, Xing and Chen, Zuhuang and Zhao, Jinzhu and Das, Sujit and Klewe, Christoph and Zhang, Lei and Lupi, Eduardo and Shafer, Padraic and others},
  journal={Advanced Functional Materials},
  volume={29},
  number={50},
  pages={1907072},
  year={2019},
  publisher={Wiley Online Library}
}

@article{yao2014electron,
  title={Electron-Beam-Induced Perovskite--Brownmillerite--Perovskite Structural Phase Transitions in Epitaxial \uppercase{L\lowercase{a}\textsubscript{2/3}S\lowercase{r}\textsubscript{1/3}M\lowercase{n}O\textsubscript{3}} Films},
  author={Yao, Lide and Majumdar, Sayani and {\"A}k{\"a}slompolo, Laura and Inkinen, Sampo and Qin, Qi Hang and van Dijken, Sebastiaan},
  journal={Advanced Materials},
  volume={26},
  number={18},
  pages={2789--2793},
  year={2014},
  publisher={Wiley Online Library}
}

@article{park2016electrochemical,
  title={Electrochemical Li topotactic reaction in layered \uppercase{S\lowercase{n}P\textsubscript{3}} for superior Li-ion batteries},
  author={Park, Jae-Wan and Park, Cheol-Min},
  journal={Scientific Reports},
  volume={6},
  number={1},
  pages={35980},
  year={2016},
  publisher={Nature Publishing Group UK London}
}

@article{takimoto2023platinum,
  title={Platinum nanosheets synthesized via topotactic reduction of single-layer platinum oxide nanosheets for electrocatalysis},
  author={Takimoto, Daisuke and Toma, Shino and Suda, Yuya and Shirokura, Tomoki and Tokura, Yuki and Fukuda, Katsutoshi and Matsumoto, Masashi and Imai, Hideto and Sugimoto, Wataru},
  journal={Nature Communications},
  volume={14},
  number={1},
  pages={19},
  year={2023},
  publisher={Nature Publishing Group UK London}
}

@article{popel2016structural,
  title={Structural effects in \uppercase{UO\textsubscript{2}} thin films irradiated with U ions},
  author={Popel, AJ and Adamska, Anna Maria and Martin, Peter George and Payton, Oliver D and Lampronti, GI and Picco, Loren and Payne, Liam and Springell, Ross and Scott, Thomas Bligh and Monnet, I and others},
  journal={Nuclear Instruments and Methods in Physics Research Section B: Beam Interactions with Materials and Atoms},
  volume={386},
  pages={8--15},
  year={2016},
  publisher={Elsevier}
}

@article{cubicciotti1952reaction,
  title={The reaction between uranium and oxygen},
  author={Cubicciotti, Daniel},
  journal={Journal of the American Chemical Society},
  volume={74},
  number={4},
  pages={1079--1081},
  year={1952},
  publisher={ACS Publications}
}

@article{rennie2018study,
  title={Study of phonons in irradiated epitaxial thin films of \uppercase{UO\textsubscript{2}}},
  author={Rennie, Sophie and Lawrence Bright, E and Darnbrough, James E and Paolasini, Luigi and Bosak, Alexei and Smith, AD and Mason, Nigel and Lander, Gerard H and Springell, Ross},
  journal={Physical Review B},
  volume={97},
  number={22},
  pages={224303},
  year={2018},
  publisher={APS}
}

@article{steeb1961elektronenbeugungs,
  title={Elektronenbeugungs-untersuchung an einkristallinen schichten von uranoxyden im bereich von \uppercase{UO\textsubscript{2}} bis \uppercase{U\textsubscript{4}O\textsubscript{9}}},
  author={Steeb, S},
  journal={Journal of Nuclear Materials},
  volume={3},
  number={2},
  pages={235--236},
  year={1961}
}

@article{ilton2011xps,
  title={\uppercase{XPS} determination of uranium oxidation states},
  author={Ilton, Eugene S and Bagus, Paul S},
  journal={Surface and Interface Analysis},
  volume={43},
  number={13},
  pages={1549--1560},
  year={2011},
  publisher={Wiley Online Library}
}

@article{vaney2022topotactic,
  title={Topotactic fluorination of intermetallics as an efficient route towards quantum materials},
  author={Vaney, Jean-Baptiste and Vignolle, Baptiste and Demourgues, Alain and Gaudin, Etienne and Durand, Etienne and Labrug{\`e}re, Christine and Bernardini, Fabio and Cano, Andr{\'e}s and Tenc{\'e}, Sophie},
  journal={Nature Communications},
  volume={13},
  number={1},
  pages={1462},
  year={2022},
  publisher={Nature Publishing Group UK London}
}

@article{lee2018redox,
  title={Redox-Driven Nanoscale Topotactic Transformations in Epitaxial \uppercase{S\lowercase{r}F\lowercase{e}\textsubscript{0.8}C\lowercase{o}\textsubscript{0.2}O\lowercase{\textsubscript{3-x}}} under Atmospheric Pressure},
  author={Lee, Joonhyuk and Ahn, Eunyoung and Seo, Yu-Seong and Kim, Younghak and Jeon, Tae-Yeol and Cho, Jinhyung and Lee, Inwon and Jeen, Hyoungjeen},
  journal={Physical Review Applied},
  volume={10},
  number={5},
  pages={054035},
  year={2018},
  publisher={APS}
}

@article{hu2022atomic,
  title={Atomic-scale observation of strain-dependent reversible topotactic transition in \uppercase{L\lowercase{a}\textsubscript{0.7}S\lowercase{r}O\textsubscript{0.3}M\lowercase{n}O\lowercase{\textsubscript{x}}} films under an ultra-high vacuum environment},
  author={Hu, Kejun and Zhang, Xinyu and Chen, Pingfan and Lin, Renju and Zhu, Jinlong and Huang, Zhen and Du, Haifeng and Song, Dongsheng and Ge, Binghui},
  journal={Materials Today Physics},
  volume={29},
  pages={100922},
  year={2022},
  publisher={Elsevier}
}

@book{bhadeshia2017geometry,
  title={Geometry of crystals, polycrystals, and phase transformations},
  author={Bhadeshia, Harshad KDH},
  year={2017},
  publisher={CRC press}
}

@book{birks2006introduction,
  title={Introduction to the high temperature oxidation of metals},
  author={Birks, Neil and Meier, Gerald H and Pettit, Frederick S},
  year={2006},
  publisher={Cambridge university press}
}

@article{springell2023review,
  title={A review of uranium-based thin films},
  author={Springell, R and Lawrence Bright, E and Chaney, DA and Harding, Lottie M and Bell, Christopher and Ward, Roger CC and Lander, Gerard H},
  journal={Advances in Physics},
  volume={71},
  number={3-4},  
  pages={1-79},
  year={2022},
  publisher={Taylor \& Francis}
}

@article{willis1978defect,
  title={The defect structure of hyper-stoichiometric uranium dioxide},
  author={Willis, BTM},
  journal={Foundations of Crystallography},
  volume={34},
  number={1},
  pages={88--90},
  year={1978},
  publisher={International Union of Crystallography}
}

@article{pijolat1997reduction,
  title={Reduction of uranium oxide U3O8 to UO2 by hydrogen},
  author={Pijolat, Mich{\`e}le and Brun, Catherine and Valdivieso, Fran{\c{c}}oise and Soustelle, Michel},
  journal={Solid State Ionics},
  volume={101},
  pages={931--935},
  year={1997},
  publisher={Elsevier}
}

@book{belle1961uranium,
  title={Uranium dioxide: properties and nuclear applications},
  author={Belle, Jack},
  volume={4},
  year={1961},
  publisher={Naval Reactors, Division of Reactor Development, US Atomic Energy Commission}
}

@article{rennie2018role,
  title={The role of crystal orientation in the dissolution of \uppercase{UO\textsubscript{2}} thin films},
  author={Rennie, S and Bright, E Lawrence and Sutcliffe, JE and Darnbrough, JE and Burrows, R and Rawle, J and Nicklin, C and Lander, GH and Springell, R},
  journal={Corrosion science},
  volume={145},
  pages={162--169},
  year={2018},
  publisher={Elsevier}
}

@article{bao2013antiferromagnetism,
  title={Antiferromagnetism in \uppercase{UO\textsubscript{2}} thin epitaxial films},
  author={Bao, Zhaoui and Springell, Ross and Walker, HC and Leiste, H and Kuebel, K and Prang, R and Nisbet, Gareth and Langridge, S and Ward, RCC and Gouder, Thomas and others},
  journal={Physical Review B},
  volume={88},
  number={13},
  pages={134426},
  year={2013},
  publisher={APS}
}

@article{strehle2012characterization,
  title={Characterization of single crystal uranium-oxide thin films grown via reactive-gas magnetron sputtering on yttria-stabilized zirconia and sapphire},
  author={Strehle, Melissa M and Heuser, Brent J and Elbakhshwan, Mohamed S and Han, Xiaochun and Gennardo, David J and Pappas, Harrison K and Ju, Hyunsu},
  journal={Thin Solid Films},
  volume={520},
  number={17},
  pages={5616--5626},
  year={2012},
  publisher={Elsevier}
}

@article{leinders2016low,
  title={Low-temperature oxidation of fine \uppercase{UO\textsubscript{2}} powders: A process of nanosized domain development},
  author={Leinders, Gregory and Pakarinen, Janne and Delville, Rémi and Cardinaels, Thomas and Binnemans, Koen and Verwerft, Marc},
  journal={Inorganic Chemistry},
  volume={55},
  number={8},
  pages={3915--3927},
  year={2016},
  publisher={ACS Publications}
}

@article{bevan1986crystal,
  title={The crystal structure of $\beta$-\uppercase{U\textsubscript{4}O}\textsubscript{9-y}},
  author={Bevan, DJM and Grey, IE and Willis, BTM},
  journal={Journal of Solid State Chemistry},
  volume={61},
  number={1},
  pages={1--7},
  year={1986},
  publisher={Elsevier}
}

@article{eriksen2012radiation,
  title={Radiation induced dissolution of \uppercase{UO\textsubscript{2}} based nuclear fuel--A critical review of predictive modelling approaches},
  author={Eriksen, Trygve E and Shoesmith, David W and Jonsson, Mats},
  journal={Journal of Nuclear Materials},
  volume={420},
  number={1-3},
  pages={409--423},
  year={2012},
  publisher={Elsevier}
}

@book{tomar2023nuclear,
  title={Nuclear Fuel Cycle},
  author={Tomar, B.S. and Rao, P.R.V. and Roy, S.B. and Panakkal, J.P. and Raj, K. and Nandakumar, A.N.},
  isbn={9789819909490},
  year={2023},
  publisher={Springer Nature Singapore}
}

@book{international2022status,
  title={Status and Trends in Spent Fuel and Radioactive Waste Management},
  author={IAEA},
  isbn={9789201305213},
  series={IAEA Nuclear Energy Series},
  year={2022},
  publisher={International Atomic Energy Agency}
}

@article{nangia2019crystal,
  title={Crystal engineering: an outlook for the future},
  author={Nangia, Ashwini K and Desiraju, Gautam R},
  journal={Angewandte Chemie International Edition},
  volume={58},
  number={13},
  pages={4100--4107},
  year={2019},
  publisher={Wiley Online Library}
}

@article{desiraju2007crystal,
  title={Crystal engineering: a holistic view},
  author={Desiraju, Gautam R},
  journal={Angewandte Chemie International Edition},
  volume={46},
  number={44},
  pages={8342--8356},
  year={2007},
  publisher={Wiley Online Library}
}

@book{wang2023nuclear,
  title={Nuclear Power Reactor Designs: From History to Advances},
  author={Wang, J. and Talabi, S. and Leon, S.B.},
  isbn={9780323999465},
  year={2023},
  publisher={Elsevier Science}
}

@article{allen1986formation,
  title={The formation of \uppercase{U\textsubscript{3}O\textsubscript{8}} on crystalline \uppercase{UO\textsubscript{2}}},
  author={Allen, GC and Tempest, PA and Tyler, JW},
  journal={Philosophical Magazine B},
  volume={54},
  number={2},
  pages={L67--L71},
  year={1986},
  publisher={Taylor \& Francis}
}

@article{allen1987oxidation,
  title={Oxidation of crystalline \uppercase{UO\textsubscript{2}} studied using X-ray photoelectron spectroscopy and X-ray diffraction},
  author={Allen, Geoffrey C and Tempest, Paul A and Tyler, Jonathan W},
  journal={Journal of the Chemical Society, Faraday Transactions 1: Physical Chemistry in Condensed Phases},
  volume={83},
  number={3},
  pages={925--935},
  year={1987},
  publisher={Royal Society of Chemistry}
}

@article{schneider2012nih,
  title={NIH Image to ImageJ: 25 years of image analysis},
  author={Schneider, Caroline A and Rasband, Wayne S and Eliceiri, Kevin W},
  journal={Nature methods},
  volume={9},
  number={7},
  pages={671--675},
  year={2012},
  publisher={Nature Publishing Group}
}

@article{esteves2017lipras,
  title={LIPRAS: Line-profile analysis software},
  author={Esteves, Giovanni and Ramos, Klarissa and Fancher, Chris M and Jones, Jacob L},
  journal={Preprint at https://www. researchgate. net/publication/316985889\_LIPRAS\_Line-Profile\_Analysis\_Software},
  year={2017}
}

@article{mceachern1998review,
  title={A review of the oxidation of uranium dioxide at temperatures below 400\textdegree{}\uppercase{C}},
  author={McEachern, Rod Joseph and Taylor, P},
  journal={Journal of Nuclear Materials},
  volume={254},
  number={2-3},
  pages={87--121},
  year={1998},
  publisher={Elsevier}
}

@article{popel2020atomic,
  title={An atomic-scale understanding of \uppercase{UO\textsubscript{2}} surface evolution during anoxic dissolution},
  author={Popel, Aleksej J and Spurgeon, Steven R and Matthews, Bethany and Olszta, Matthew J and Tan, Beng Thye and Gouder, Thomas and Eloirdi, Rachel and Buck, Edgar C and Farnan, Ian},
  journal={ACS applied materials \& interfaces},
  volume={12},
  number={35},
  pages={39781--39786},
  year={2020},
  publisher={ACS Publications}
}

@article{meng2023topotactic,
  title={Topotactic Transition: A Promising Opportunity for Creating New Oxides},
  author={Meng, Ziang and Yan, Han and Qin, Peixin and Zhou, Xiaorong and Wang, Xiaoning and Chen, Hongyu and Liu, Li and Liu, Zhiqi},
  journal={Advanced Functional Materials},
  volume={33},
  number={46},
  pages={2305225},
  year={2023},
  publisher={Wiley Online Library}
}

@article{vallejo2022advances,
  title={Advances in actinide thin films: synthesis, properties, and future directions},
  author={Vallejo, Kevin D and Kabir, Firoza and Poudel, Narayan and Marianetti, Chris A and Hurley, David H and Simmonds, Paul J and Dennett, Cody A and Gofryk, Krzysztof},
  journal={Reports on Progress in Physics},
  volume={85},
  number={12},
  pages={123101},
  year={2022},
  publisher={IOP Publishing}
}

@article{middleburgh2021structure,
  title={Structure and properties of amorphous uranium dioxide},
  author={Middleburgh, Simon C and Lee, William E and Rushton, Michael JD},
  journal={Acta Materialia},
  volume={202},
  pages={366--375},
  year={2021},
  publisher={Elsevier}
}

@article{spurgeon2019nanoscale,
  title={Nanoscale oxygen defect gradients in \uppercase{UO\lowercase{\textsubscript{2+x}}} surfaces},
  author={Spurgeon, Steven R and Sassi, Michel and Ophus, Colin and Stubbs, Joanne E and Ilton, Eugene S and Buck, Edgar C},
  journal={Proceedings of the national academy of sciences},
  volume={116},
  number={35},
  pages={17181--17186},
  year={2019},
  publisher={National Acad Sciences}
}

@article{desgranges2009neutron,
  title={Neutron diffraction study of the in situ oxidation of \uppercase{UO\textsubscript{2}}},
  author={Desgranges, Lionel and Baldinozzi, Gianguido and Rousseau, Gurvan and Niepce, Jean-Claude and Calvarin, Gilbert},
  journal={Inorganic chemistry},
  volume={48},
  number={16},
  pages={7585--7592},
  year={2009},
  publisher={ACS Publications}
}

@article{idriss2010surface,
  title={Surface reactions of uranium oxide powder, thin films and single crystals},
  author={Idriss, Hicham},
  journal={Surface Science Reports},
  volume={65},
  number={3},
  pages={67--109},
  year={2010},
  publisher={Elsevier}
}

@article{aronson1957kinetic,
  title={Kinetic study of the oxidation of uranium dioxide},
  author={Aronson, S and Roof Jr, RB and Belle, J},
  journal={The Journal of Chemical Physics},
  volume={27},
  number={1},
  pages={137--144},
  year={1957},
  publisher={American Institute of Physics}
}

@article{hoekstra1961low,
  title={The low temperature oxidation of \uppercase{UO\textsubscript{2}} and \uppercase{U\textsubscript{4}O\textsubscript{9}}},
  author={Hoekstra, HR and Santoro, A and Siegel, S},
  journal={Journal of Inorganic and Nuclear Chemistry},
  volume={18},
  pages={166--178},
  year={1961},
  publisher={Elsevier}
}

@article{westrum1962triuranium,
  title={Triuranium heptaoxides: Heat capacities and thermodynamic properties of $\alpha$-and $\beta$-\uppercase{U\textsubscript{3}O\textsubscript{7}} from 5 to 350 \uppercase{K}},
  author={Westrum Jr, Edgar F and Gr{\o}nvold, F},
  journal={Journal of Physics and Chemistry of Solids},
  volume={23},
  number={1-2},
  pages={39--53},
  year={1962},
  publisher={Elsevier}
}

@article{rousseau2006detailed,
  title={A detailed study of \uppercase{UO\textsubscript{2}} to \uppercase{U\textsubscript{3}O\textsubscript{8}} oxidation phases and the associated rate-limiting steps},
  author={Rousseau, G and Desgranges, L and Charlot, F and Millot, N and Ni{\`e}pce, JC and Pijolat, M and Valdivieso, F and Baldinozzi, Gianguido and B{\'e}rar, JF},
  journal={Journal of nuclear materials},
  volume={355},
  number={1-3},
  pages={10--20},
  year={2006},
  publisher={Elsevier}
}

@article{desgranges2011refinement,
  title={Refinement of the $\alpha$-\uppercase{U\textsubscript{4}O\textsubscript{9}} crystalline structure: New insight into the \uppercase{U\textsubscript{4}O\textsubscript{9}}→ \uppercase{U\textsubscript{3}O\textsubscript{8}} transformation},
  author={Desgranges, Lionel and Baldinozzi, Gianguido and Simeone, David and Fischer, HE},
  journal={Inorganic Chemistry},
  volume={50},
  number={13},
  pages={6146--6151},
  year={2011},
  publisher={ACS Publications}
}

@article{walker1965oxidation,
  title={The oxidation of uranium dioxides},
  author={Walker, DEY},
  journal={Journal of Applied Chemistry},
  volume={15},
  number={3},
  pages={128--135},
  year={1965},
  publisher={Wiley Online Library}
}

@article{allen1995mechanism,
  title={A mechanism for the \uppercase{UO\textsubscript{2}} to $\alpha$-\uppercase{U\textsubscript{3}O\textsubscript{8}} phase transformation},
  author={Allen, GC and Holmes, NR},
  journal={Journal of Nuclear Materials},
  volume={223},
  number={3},
  pages={231--237},
  year={1995},
  publisher={Elsevier}
}

@article{ackermann1977thermal,
  title={Thermal expansion and phase transformations of the \uppercase{U\textsubscript{3}O\textsubscript{8}}- z phase in air},
  author={Ackermann, RJ and Chang, AT and Sorrell, Charles A},
  journal={Journal of Inorganic and Nuclear Chemistry},
  volume={39},
  number={1},
  pages={75--85},
  year={1977},
  publisher={Elsevier}
}

@article{song1994formation,
  title={Formation of columnar \uppercase{U\textsubscript{3}O\textsubscript{8}} grains on the oxidation of \uppercase{UO\textsubscript{2}} pellets in air at 900°\uppercase{C}},
  author={Song, Kun Woo and Yang, Myung Seung},
  journal={Journal of nuclear materials},
  volume={209},
  number={3},
  pages={270--273},
  year={1994},
  publisher={Elsevier}
}

@article{taylor1992early,
  title={The early stages of \uppercase{U\textsubscript{3}O\textsubscript{8}} formation on unirradiated CANDU \uppercase{UO\textsubscript{2}} fuel oxidized in air at 200--300°\uppercase{C}},
  author={Taylor, Peter and Wood, Donald D and Duclos, A Michael},
  journal={Journal of nuclear materials},
  volume={189},
  number={1},
  pages={116--123},
  year={1992},
  publisher={Elsevier}
}

@article{bae1994oxidation,
  title={Oxidation behavior of unirradiated \uppercase{UO\textsubscript{2}} pellets},
  author={Bae, KK and Kim, BG and Lee, YW and Yang, MS and Park, HS},
  journal={Journal of nuclear materials},
  volume={209},
  number={3},
  pages={274--279},
  year={1994},
  publisher={Elsevier}
}

@article{quemard2009origin,
  title={On the origin of the sigmoid shape in the \uppercase{UO\textsubscript{2}} oxidation weight gain curves},
  author={Qu{\'e}mard, Ludovic and Desgranges, Lionel and Bouineau, Vincent and Pijolat, Mich{\`e}le and Baldinozzi, GUIDO and Millot, Nadine and Ni{\`e}pce, Jean-Claude and Poulesquen, ARNAUD},
  journal={Journal of the European Ceramic Society},
  volume={29},
  number={13},
  pages={2791--2798},
  year={2009},
  publisher={Elsevier}
}

@phdthesis{wasik2021oxidation,
  title={Oxidation of Uranium Dioxide},
  author={Wasik, Jacek M},
  year={2021},
  school={University of Bristol}
}

\end{document}